\documentclass{raa}
\usepackage{graphicx,times}
\usepackage{natbib}
\usepackage{amssymb,amsmath}
\bibpunct{(}{)}{;}{a}{}{,}
\usepackage{longtable}

\usepackage[a4paper=true,dvipdfm=true,pagebackref=true]{hyperref}
\hypersetup{pdftitle = The title of my PDF, pdfauthor = My name, pdfsubject= The subject, pdfkeywords = keyword1 keyword2 keyword3}
\hypersetup{colorlinks = true, linkcolor = green, anchorcolor = red, citecolor = blue, filecolor = red, pagecolor = red, urlcolor = red}

\begin{document}

\title{Photometric Investigation of the K-type Extreme-Shallow Contact Binary V1799 Orion
$^*$
\footnotetext{\small $*$ Supported by the National Natural Science Foundation of China.}
}

 \volnopage{ {\bf 2012} Vol.\ {\bf X} No. {\bf XX}, 000--000}
   \setcounter{page}{1}

\author{Nian-Ping Liu\inst{1,2,3}, Sheng-Bang Qian\inst{1,2,3}, Wen-Ping Liao
 \inst{1,2},  Jia-Jia He\inst{1,2}, Er-Gang Zhao\inst{1,2}, Liang Liu\inst{1,2}
   }

   \institute{ Yunnan Observatories, Chinese Academy of Sciences, Kunming, 650011
China; {\it lnp@ynao.ac.cn}\\
    \and
    Key Laboratory for the Structure and Evolution of Celestial Objects, Chinese Academy of
Sciences, 650011 Kunming, China\\
    \and
    University of Chinese Academy of Sciences, Beijing 100049, China\\
\vs \no
   {\small Received ; accepted }
}

\abstract{New multi-color light curves of the very short period K-type eclipsing binary
V1799 Ori were obtained and analyzed with the W-D code. The photometric solutions
reveal that the system is a W-type shallow-contact binary with a mass ratio of
$q=1.335(\pm0.005)$ and a degree of contact about $f = 3.5(\pm1.1)\%$. In general,
the results are in good agreement with which is reported by Samec. The remarkable O'Connell
effects in the light curves are well explained by employing star spots on the binary
surface, which confirms that the system is active at present. Several new times of
light minimum were obtained. All the available times of light minimum were collected,
along with the recalculated and new obtained. Applying a least-squares
method to the constructed O-C diagram, a new ephemeris was derived for V1799 Ori.
The orbital period is found to show a continuous weak increase at a rate of
$1.8(\pm0.6)\times10^{-8}$ days$\cdot$yr$^{-1}$. The extreme-shallow contact,
together with the period increase, suggests that the binary may be at a critical
stage predicted by the TRO theory.
}

\authorrunning{N.-P. Liu et al. }            
\titlerunning{Photometric Investigation of the K-type Extreme-Shallow Contact Binary V1799 Orion}  
\maketitle

%
\section{Introduction}           
\label{sect:intro}
The K-type short-period contact binaries, especially those with periods shorter than 0.3 days,
are most probably main-sequence objects and mostly convective \citep{Bradstreet85}.
They belong to W UMa type binaries which both components share common envelope
that lying between the inner and outer critical Roche-lobe surfaces.
W UMa type binaries have been divided into two sub-groups: A-type and W-type
according to \citet{Binnendijk70}. Generally speaking, the A-type systems are more
likely to have earlier spectral type than the W-type systems \citep{Rucinski74}.
So it is supposed that a large amount of K-type short-period contact binaries
may be W-type systems. The majority of W-type contact binaries show shallow contact
characteristics (\citealt{ZhuL10}; He 2009 PhD). Therefore, they are good targets
for testing the thermal relaxation oscillation theory (TRO theory; eg.,
\citealt{Lucy76,Flannery76,Robertson77})

V1799 Orion, or V1799 Ori (=GSC 00096-00175, $\alpha_{2000.0}$ = $04^{h}47^{m}18^{s}.19$ and
$\delta_{2000.0}$ = $+06^{\circ}40'56''.1$) is a very short period (with period shorter than 0.3 days)
eclipsing binary. It was first suspected to show variability by \citet{Hanley40Shap}
(the cross identification is HV 10397). However, it had been neglected for a long time until
ROSTE survey \citep{Akerlof00} rediscovered it to be a eclipsing binary with EW type light curves
(Khruslove \citealt{Khrus1ov06}) (the cross identification is NSV 1719).
The following ephemeris was reported (Khruslove \citealt{Khrus1ov06}):
\begin{equation}
\mathrm{Min.I (HJD)} = 2451524.829+0^{\mathrm{d}}.29031\times E.
\end{equation}
It was then monitored many times by several researchers such as
\citet[][etc.]{Diethelm09,Diethelm10} and \citet{Nelson10}.
Recently, The O-C gateway (http://var.astro.cz/ocgate/) gave out a more exact ephemeris,
\begin{equation}\label{eq:ephe}
\mathrm{Min.I (HJD)} = 2451524.829+0^{\mathrm{d}}.290304\times E.
\end{equation}

The first photometric analysis of V1799 Ori was given by \cite{Samec10} as a student/professional
collaborative program. Their photometric solutions depicted V1799 Ori as a W-type,
extremely shallow-contact, eclipsing binary. The degree of contact factor $3\,\%$
is rather exceptional. Their solutions uncovered two hot spots on the components,
which indicates the system is quite active at present.

In this paper, we present the analysis of newly obtained complete CCD light curves and
compare the results with those given by Samec et al. Especially, we carry out a more
exact period investigation of this system, using all the available times of light minimum.


\section{Observations}
\label{sect:Obs}
\begin{figure}
\begin{center}
\includegraphics[width=9.5cm]{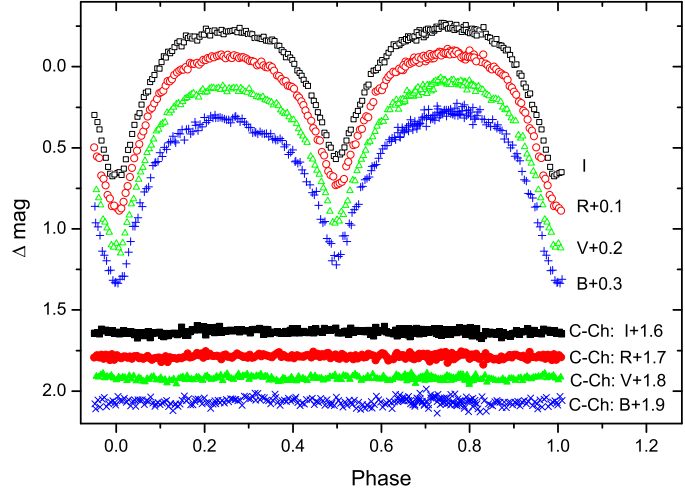}
\caption{Multi-color light curves of V1799 Ori obtained in 2012.}\label{fig:lc}
\end{center}
\end{figure}

New CCD photometric observations of V1799 Ori in BV(RI)$_{\rm c}$ bands were carried
out on December 15, 2012 with the 1.0-m reflecting telescope at Yunnan Observatory of
Chinese Academy of Sciences. An Andor DW436 2K CCD camera equiped in Cassegrain-focus was used.
A standard Johnson-Cousins-Bessel filter system built on the primary focus \citep{ZhouA09}
was used. The effective field of view was $7.3' \times 7.3'$. The integration times
for each image were 40 s, 30 s, 15 s and 10 s in B, V, R, and I bands, respectively.
The stars 2MASS J04472543+0641432 and 2MASS J04472765+0638427 near the target were
chosen as the comparison and the check stars, respectively. Their brightness and color are
similar to that of the variable. The raw images were reduced with PHOT
(magnitudes measurements for a list of stars) task of the aperture photometry package IRAF.
Complete BVRI light curves were obtained and the original data of the light curves are
listed in Tables \ref{tabA1}-\ref{tabA4}. The phased light curves are displayed in Figure \ref{fig:lc}.
The phases were calculated with respect to the following ephemeris,
\begin{equation}
\mathrm{Min.I(HJD)} = 2456277.10109+0^{\mathrm{d}}.290304\times{E}.
\end{equation}
The initial epoch in this equation is the averaged time of min I (see Table
\ref{tabnewm}, data with Hel. JD $2456277\sim$ ).
Some information about the light curves are listed in Table \ref{tabsumlc} to describe their
main features. By the way, more CCD observations of V1799 Ori were made on
two nights in 2013 December with the same telescope. These data were obtained
around the eclipsing times and were used to determine the times of light minimum
(see Table \ref{tabnewm}).

\begin{table}
\begin{center}
\caption{Summary of the multi-color light curves}\label{tabsumlc}
\begin{tabular}{cccc}\hline\hline
Wave Band  & min.I-min.II & max.(0.25)-max.(0.75) &  min.I-max.(0.75)\\
           &   (mag.)     &   (mag.)              &   (mag.)     \\\hline
   B       & 0.15         & 0.05                  &  1.08  \\
   V       & 0.18         & 0.04                  &  1.03  \\
   R       & 0.13         & 0.03                  &  0.98  \\
   I       & 0.10         & 0.03                  &  0.93  \\
\hline
\end{tabular}
\end{center}
\end{table}

The light curves displayed in Figure \ref{fig:lc} show remarkable O'Connell effects
(different heights of light maximum, see \citealt{OConnell51a, OConnell51b}, \citealt{Milone68}),
which we should taken into account in deriving photometric solutions.
The two deep eclipses of the light curves will help us to derive more
reliable parameters of the binary. Using a least-square parabolic fitting method,
the new times of CCD light minimum were determined and are listed in Table \ref{tabnewm}.

\begin{table}
\begin{center}
\begin{minipage}[]{100mm}
\caption{New times of CCD light minimum.}\label{tabnewm}\end{minipage}
\begin{tabular}{ccccc}\hline
\hline
JD (Hel.)    & Error (days)   &  Min. & Filter  & NA \\\hline
     2456277.10131 & 0.00033  &  I    & B       & 35 \\
     2456277.10091 & 0.00030  &  I    & I       & 29 \\
     2456277.10110 & 0.00032  &  I    & R       & 34 \\
     2456277.10105 & 0.00030  &  I    & V       & 30 \\
     2456277.24598 & 0.00043  &  II   & B       & 36 \\
     2456277.24612 & 0.00031  &  II   & I       & 29 \\
     2456277.24600 & 0.00034  &  II   & R       & 29 \\
     2456277.24648 & 0.00033  &  II   & V       & 30 \\
     2456645.20608 & 0.00013  &  I    & N       & 54 \\
     2456657.25337 & 0.00012  &  II   & I       & 32 \\
     2456657.25374 & 0.00016  &  II   & N       & 32 \\
\hline
\end{tabular}
\end{center}
\tablecomments{\textwidth}{NA is the total
number of data used to determine the times of light minimum.}
\end{table}

\section{Orbital period investigation of V1799 Ori}
Times of light minimum are very useful material for the orbital period study of eclipsing
binaries. Therefore, we collected all the available times of light minimum
in the help of O-C gateway\footnote{http://var.astro.cz/ocgate/}.
They are listed in Table \ref{tabmint}. Several times of light minimum
from the INTEGRAL Optical Monitoring Camera \citep{AlfonGarz12} Archive and
ROTSE-I \citep{Akerlof00} are also listed. They are recalculated by folding the original data
according to the period and applying a least-square parabolic fitting method
to the eclipsing part of the folded light curves. The resulted times
are actually the average times of light minimum for each group of data.
The HJD times of the light minimum determined from the light curves observed
by \citet{Samec10} were reprocessed according to their observation time.

\begin{table}
\caption{Collection of published or reprocessed times of light minimum}\label{tabmint}
\begin{center}
\small
\begin{tabular}{llrccccl}\hline
\hline
JD. Hel.         & Err.      &Epoch     & O-C         & Min  & Method  &   Ref. & Notes \\
2,400,000+       & (days)    &          & (days)      &      &         &        &   \\
\hline
51524.10538      & 4.8E-4    & -2.5      &  0.00214  & II    &  ccd  &  (1)   & reproc   \\
51526.28195      & 3.5E-4    &  5        &  0.00143  & I     &  ccd  &  (1)   & reproc   \\
53591.5061       &  0.003    &  7119.0   &  0.00295  & I     &  V    &  (2)   & newp, unused\\
53591.6487       &  0.002    &  7119.5   &  0.00034  & II    &  V    &  (2)   & newp, unused\\
54014.0306       &  0.003    &  8574.5   & -0.01008  & II    &  V    &  (2)   & newp, unused\\
54530.3413       &  0.001    &  10353.0  & -0.00502  & I     &  V    &  (2)   & newp       \\
54756.9226       &   9E-4    &  11133.5  & -0.00598  & II    &  B    &  (3)   & \\
54821.66301      & 1.6E-4    &  11356.5  & -0.00337  & II    &  BVRI &  (4)   & reproc   \\
54821.80835      & 2.9E-4    &  11357.0  & -0.00318  & I     &  BVRI &  (4)   & reproc   \\
54822.67936      & 2.1E-4    &  11360.0  & -0.00308  & I     &  BVRI &  (4)   & reproc   \\
54822.82433      & 4.2E-4    &  11360.5  & -0.00326  & II    &  VRI  &  (4)   & reproc   \\
54824.71134      & 1.4E-4    &  11367.0  & -0.00323  & I     &  BVRI &  (4)   & reproc   \\
54824.85607      & 7.1E-4    &  11367.5  & -0.00365  & II    &  V    &  (4)   & reproc   \\
54827.76016      & 2.2E-4    &  11377.5  & -0.00260  & II    &  BVR  &  (4)   & reproc   \\
54827.90541      & 1.5E-4    &  11378.0  & -0.00250  & I     &  BVR  &  (4)   & reproc   \\
55113.8532       &   5E-4    &  12363.0  & -0.00415  & I     &  ccd  &  (5)   & \\
55135.7727       &   4E-4    &  12438.5  & -0.00260  & II    &  V    &  (6)   & \\
55135.9187       &   5E-4    &  12439.0  & -0.00176  & I     &  V    &  (6)   & \\
55506.9248       &   3E-4    &  13717.0  & -0.00417  & I     &  V    &  (7)   & \\
55937.5900       &  0.003    &  15200.5  & -0.00495  & II    &  V    &  (8)   & unused  \\
55937.7353       &   8E-4    &  15201.0  & -0.00480  & I     &  V    &  (8)   & \\
56237.9093       &   8E-4    &  16235.0  & -0.00514  & I     &  V    &  (9)   & \\
\hline
\end{tabular}
\end{center}
\tablecomments{\textwidth}{ROTSE-I data can be retrieved in the NSVS database (http://skydot.lanl.gov/nsvs/nsvs.php).
newp = Based on data from the OMC Archive at CAB (INTA-CSIC), pre-processed by ISDC, times of light
minimum are newly calculated; reproc = data were reprocessed. \\}
\tablerefs{\textwidth}{(1)~Khrus1ov \citealt{Khrus1ov06}; (2) present paper; (3)\citealt{Diethelm09};
(4)\citealt{Samec10}; (5)\citealt{Nelson10}; (6) \citealt{Diethelm10}; (7) \citealt{Diethelm11};
(8) \citealt{Diethelm12}; (9) \citealt{Diethelm13}.}
\end{table}

\begin{figure}
\begin{center}
\includegraphics[width=10cm]{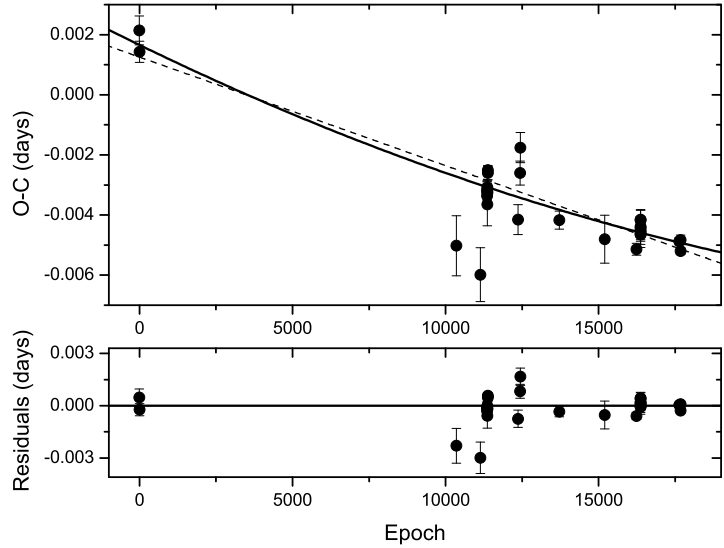}
\caption{ Upper panel: the O-C diagram calculated using the linear ephemeris in Equation \ref{eq:ephe}.
The solid line refers to the upward parabolic variation, while the the dash line refers to
a linear correction for the ephemeris.
Low panel: the residuals after the parabolic variation was removed.
}\label{fig:oc}
\end{center}
\end{figure}

The collection of times of light minimum from literature or from reprocessed data
are listed in Table \ref{tabmint}. The O-C values  calculated with respect to
Equation (\ref{eq:ephe}) are shown in Figure \ref{fig:oc} along with the epochs
(data with errors larger than 0.001 days were not used).
Seen from upper panel of Figure \ref{fig:oc}, the general trend of the O-C shows
a upward parabolic change without overlapping an obvious cyclic oscillation.
By using a weighted least square method, a new ephemeris was derived to be
\begin{equation}\label{eq:newephq}
\begin{array}{ll}
\mathrm{Min.I (HJD)}=&2451524.8307(\pm0.0003)+0^{\mathrm{d}}.29030351(\pm0.00000005)\times E \\
          &+7^{\mathrm{d}}.0(\pm2.4)\times10^{-12}\times E^2.
\end{array}
\end{equation}

The quadratic term in Equation (\ref{eq:newephq}) denotes a continuous period increase
at a rate of ${\rm d}P/{\rm d}t = 1.8(\pm0.6)\times10^{-8}$ days$\cdot$yr$^{-1}$.
The residuals are also displayed in the lower panel of Figure \ref{fig:oc}.
As shown in this figure, a few data point are not fitted well. It may be caused by
occasional factors which would impact the shape of the light curves and therefore
the determination of times of light minimum, e.g.
the occurrence of solar-like activities and the unstable weather conditions.
Nevertheless, the general trend of O-C data is well described by the upward
parabolic fit.

By the way, a revised linear ephemeris was calculated to be
\begin{equation}\label{eq:leph}
\mathrm{min. I (HJD)} = 2451524.8303(\pm0.0002) + 0^{\mathrm{d}}.29030364(\pm0.00000002)\times E,
\end{equation}
which can be used as a preliminary prediction of times of light minimum.

\section{Photometric solutions with the W-D method}
The light curves displayed in Figure \ref{fig:lc} show EW-type variation and deep eclipsing feature in both minimum.
We can predict a high orbital inclination of the binary system, which enables a reliable
photometric parameter determination. To understand the geometrical structure and evolutionary state,
the multi-color light curves were analyzed with the W-D method \citep{WilsonR71Dev, WilsonR79,
WilsonR90, WilsonR94, Wilson03van}. The temperature for star 1 (the star eclipsed at primary light minimum)
was  fixed as $T_1 = 5000$ K according to \citet{Samec10}. This is reasonable because the 2MASS color index
$J - H = 0.448 $ and $H - K = 0.089$ given in VizieR database correspond to a
K0-K2 spectral type according to Allen's table \citep{Cox00}. The component eclipsed at
the primary light minimum are usually the hotter one so that we adopted the
same temperature as Samec chose. The gravity-darkening coefficients were assumed as
$g_{1}=g_{2}=0.32$ \citep{LucyL67} and the bolometric albedo as $A_{1}=A_{2}=0.5$ \citep{Rucinski69}.
Bolometric and bandpass square-root limb-darkening parameters are taken from \cite{VanHamme93}.
Mode 3 (contact model )was assumed during process of calculation. The adjustable parameters were:
the orbital inclination $i$; the mean temperature of star 2, $T_2$; the monochromatic
luminosity of star 1, $L_{1B}$, $L_{1V}$, $L_{1R}$ and $L_{1I}$;
and the dimensionless potential ($\Omega_{1}=\Omega_{2}$ for mode 3).

\begin{figure}
\begin{center}
\includegraphics[angle=0,scale=0.6]{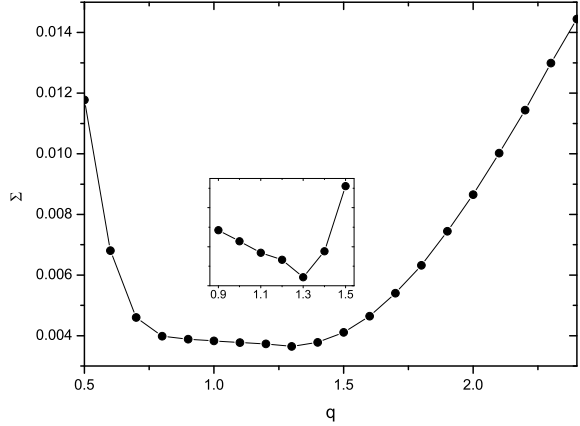}
\caption{$\Sigma$ - $q$ curves for V1799 Ori. Insert: a enlarged figure to show small part.}
\label{fig:qser}
\end{center}
\end{figure}

A q-search method was used in order to get the initial input parameters. A series
of mass ratios ranging from less than 0.3 to larger than 3 were assumed as trial values.
Calculation were carried out with these mass ratios. The resulted sum of weighted square
deviations ($\Sigma w_{i}(O - C)_{i}^{2}$) along with mass ratios are plotted in Figure
\ref{fig:qser}. The smallest $\Sigma$ was achieved at $q = 1.3$ ($q=M_{2}/M_{1}$).
Then, we treated $q$ as an an adjustable parameter and chose $q = 1.3$ as the initial value.
we treated $q$ as an an adjustable parameter and chose $q = 1.3$ as the initial value and finally
a set of converged solutions were obtained.

\begin{figure}
\begin{center}
\includegraphics[width=14cm]{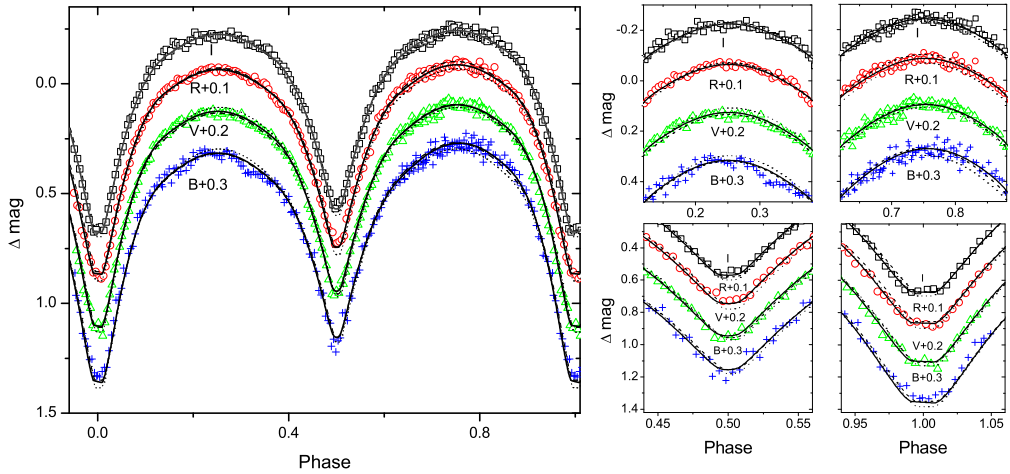}
\caption{Observed and theoretical light curves (lines) of V1799 Ori. Different colors of
symbols stand for different bands observations. The solid line, dash line and dotted line
denote theoretical light curves calculated with hot spot model 4, cool spot model 1 and
non-spot model 0, respectively. Left panel: the complete light curves. Right panels:
the enlarged small part of the left panel at minimum and maximum phases.
}\label{fig:wd}
\end{center}
\end{figure}

\begin{figure}
\begin{center}
\includegraphics[width=11cm]{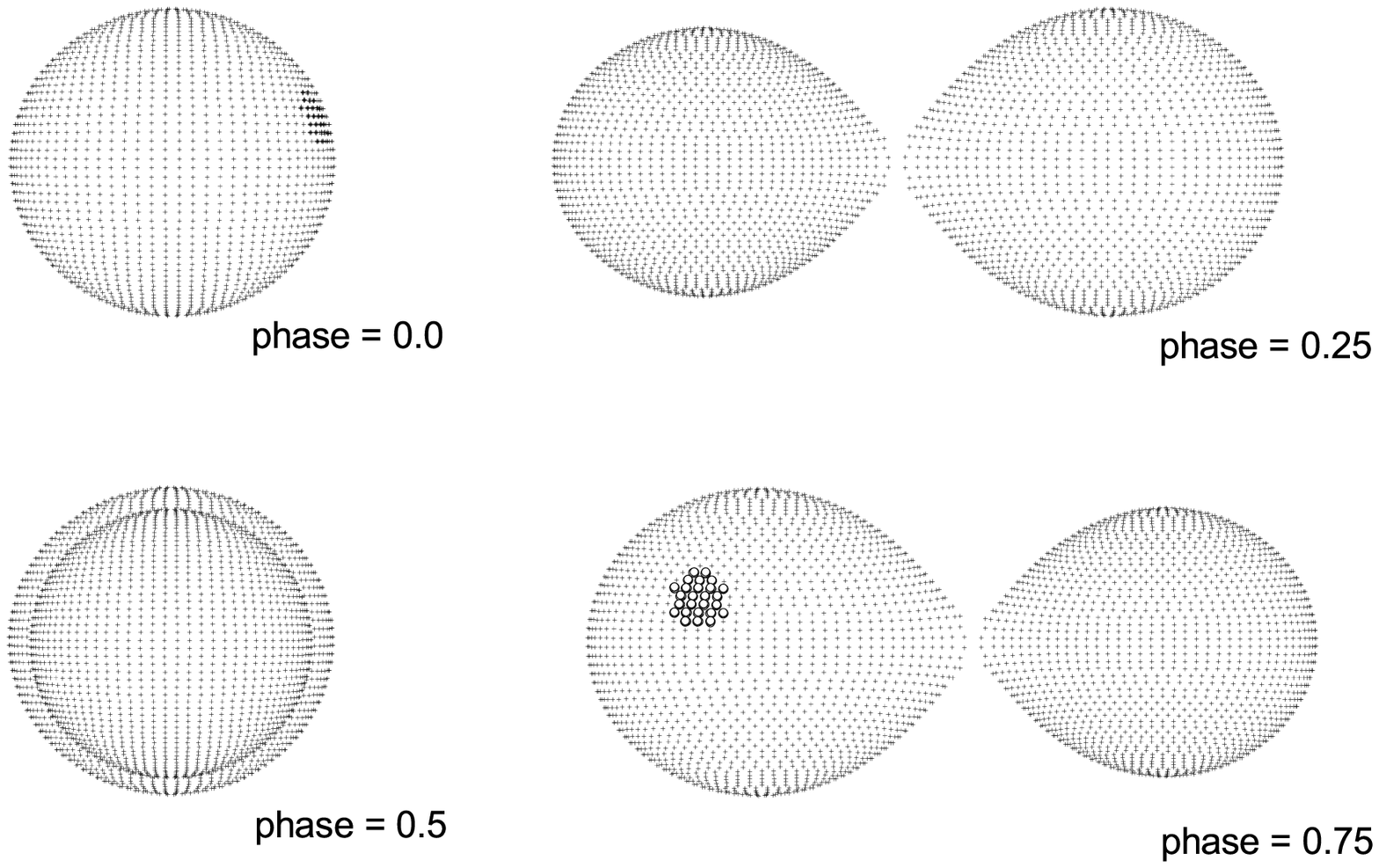}
\caption{Geometrical structure of V1799 Ori at phases 0.0, 0.25, 0.5, and 0.75 respectively.
}\label{fig:geo}
\end{center}
\end{figure}

The obtained photometric solutions are listed in Table \ref{tabsolu} (see column 2).
However, the theoretical light curves obtained does not fit the observations very well because of
the obvious asymmetry in the light curves. This is the so-called O'Connell effects.
Therefore, we tried to employ star spots in the model. It should
be noted that the spots would be cool spots, just like sunspots, or would be hot spots, which are
hotter areas on the components. Since the masses of the components are close
to each other ($q=M_{2}/M_{1}\sim1.3$), both of them may have the possibility of producing spots.
To simplify the problem, we tried adding single spot on the binary system.
Together with the non-spot model, there are five scenarios to compare, which are listed in
Table \ref{tabmode}. The values in column 2 are the sum of weighted squared residuals
for the best solution of each scenario. It could be seen that the sum of squared residuals of spotted models are
remarkably smaller than that of Non-spot model. When finding the converged solutions, we noticed that
these two kinds of cool spot models have the similar parameters. The same situation was found with those two
kinds of hot spot models. Thus we adopted model 1 and model 4 to give detailed description.
The parameters of model 0 (non-spot), model 1 (cool spot) and model 4 (hot spot)
are listed in Table \ref{tabsolu}. The corresponding theoretical light curves
of these models are plotted in Figure \ref{fig:wd}. We can see in this figure that the observed
light curves are more perfectly fitted by the hot spot model which reveals that
V1799 Ori is a W-type shallow contact binary system with a degree of contact
of $f=(\Omega-\Omega_{\mathrm{in}})/(\Omega_{\mathrm{out}}-\Omega_{\mathrm{in}})=3.5\%\pm1.1\%$.
The geometrical structure of the binary is shown in Figure \ref{fig:geo}.

\begin{table}
\begin{center}
\caption{Comparison of different spot scenarios.}\label{tabmode}
\begin{tabular}{ccc}\hline
\hline
Model &Spot type      &  $\Sigma{w_{i}(O-C)_{i}^2}\times10^{3}$ \\\hline
0  &  Non-spot        &  2.854  \\
1  &  cool spot on C1 &  2.191  \\
2  &  cool spot on C2 &  2.255  \\
3  &  hot spot on C1  &  1.935  \\
4  &  hot spot on C2  &  1.852  \\
\hline
\end{tabular}
\end{center}
\scriptsize{Notes. C1 and C2 represent component star 1 and 2 respectively.}
\end{table}

\begin{table}
\caption{Photometric Solutions for V1799 Ori from different spot scenarios}\label{tabsolu}
\begin{center}
\small
\begin{tabular}{lccc}\hline
\hline
Parameters        & Model 0         &  Model 1             & Model 4  \\
                  & Non-spot        &  Cool spot           & Hot spot \\
\hline
$q$ ($M_2/M_1$ )          & $1.300\pm0.003$    & $1.300\pm0.007$    & $1.335\pm0.005$ \\
$\Omega_{in}$             & 4.2233             & 4.2233             & 4.2771\\
$\Omega_{out}$            & 3.6571             & 3.6571             & 3.7087\\
$T_{2}$ (K)               & $4805\pm5$         & $4801\pm4$         & $4781\pm4$   \\
$i(^{\circ})$             & $89.8\pm0.9$       & $89.8\pm0.8$       &$89.7\pm0.8$\\
$L_{1}/(L_{1}+L_{2})$ (B) &$0.5176\pm0.0022$   & $0.5196\pm0.0019$  &$0.5206\pm0.0017$\\
$L_{1}/(L_{1}+L_{2}$) (V) &$0.5022\pm0.0018$   & $0.5038\pm0.0016$  &$0.5033\pm0.0014$\\
$L_{1}/(L_{1}+L_{2})$ (R) &$0.4900\pm0.0015$   & $0.4914\pm0.0015$  &$0.4895\pm0.0012$\\
$L_{1}/(L_{1}+L_{2}$) (I) &$0.4823\pm0.0013$   & $0.4835\pm0.0014$  &$0.4809\pm0.0011$\\
$\Omega_{1}=\Omega_{2}$   &$4.1763\pm0.0066$   & $4.1695\pm0.0119$  &$4.2572\pm0.0064$\\
$r_{1}(pole)$              &$0.3395\pm0.0008$  & $0.3403\pm0.0015$  &$0.3343\pm0.0008$\\
$r_{1}(side)$              &$0.3564\pm0.0010$  & $0.3573\pm0.0018$  &$0.3504\pm0.0010$\\
$r_{1}(back)$              &$0.3910\pm0.0015$  & $0.3924\pm0.0029$  &$0.3833\pm0.0016$\\
$r_{2}(pole)$              &$0.3832\pm0.0008$  & $0.3839\pm0.0014$  &$0.3824\pm0.0008$\\
$r_{2}(side)$              &$0.4048\pm0.0010$  & $0.4058\pm0.0017$  &$0.4037\pm0.0010$\\
$r_{2}(back)$              &$0.4372\pm0.0014$  & $0.4385\pm0.0025$  &$0.4345\pm0.0014$\\
$f(\%)$                    & $8.3\pm1.2$       & $9.5\pm2.1$        &$3.5\pm1.1$    \\
$\theta_{s}(^{\circ})$     & --                &   83.1(trial)    &  74.5(trial)  \\
$\psi_{s}(^{\circ})$       & --                &  $262\pm17$        & $246\pm10$    \\
$r_{s}(^{\circ})$          & --                &  $16.6\pm6.1$      & $10.8\pm5.9$  \\
$T_{s}/T_{\ast}$           & --                &  $0.88\pm0.11$     & $1.20\pm0.11$ \\
$\Sigma{w_{i}(O-C)_{i}^2}$ & 0.002854          & 0.002191           &0.001852\\
\hline
\end{tabular}
\end{center}
\end{table}

\section{Conclusion and Discussion}
\label{sect:discussion}
The deep eclipses in both primary and secondary light minimum which denotes high inclination of the
binary system help us to give reliable solutions of the system.
Based on the complete BV(RI)$_{\rm c}$ light curves, the photometric solutions for V1799 Ori
were carefully derived. We found that V1799 Ori is a extreme-shallow contact binary system with
a mass ratio of $q = M_{2}/M_{1} = 1.335\pm0.005$ and a degree of contact about $f = 3.5\%\pm1.1\%$.
It is a W-type system of which the less massive component is about 220 K hotter than the
more massive one. The asymmetric light curves can be well modeled by employing a hot spot on the
primary. Assuming that the primary component is a K0 type main-sequence star, its mass was
roughly estimated to be $M_{2}=0.80$ M$_{\odot}$ \citep{Cox00}. Then the mass the other component was
estimated to be $M_{1}=0.60$ M$_{\odot}$ by using the derived mass ratio.

The photometric solution is in good agreement with that given by \cite{Samec10}
except for the configuration of the star spots. This can be explained by the activity of the components.
The solar-like activities on the surface of the binaries are expected to change with time,
which would cause different patterns of star spots. When examining the light curves
obtained by Samec et al., we found their shapes are slightly different from ours.
The difference between the two maximum in their light curves are about 0.03 mag. (B,V),
0.02 mag (R), and 0.01 mag. (I), which are distinctively smaller than that in the newly
obtained light curves (see Table \ref{tabmode}). This further confirms that
the binary is highly active at present.

Based on the analysis of the O-C diagram (Figure \ref{fig:oc}), a general trend of
long-term period increase at a rate of $1.8(\pm0.6)\times10^{-8}$ days$\cdot$yr$^{-1}$
was derived.  Although the period increase is insignificant (close to the error),
it is no doubt a small one compared with other shallow contact binaries
(see Table 7 in the paper \citealt{ZhuL10}). This is in accord with the exceptional low
degree of contact, which also agrees that the period-increased system usually have
a lower degree of contact \citep{ZhuL10}. The long-term period increase, together with
the exceptional low degree of contact, suggests that the binary may be at a critical
stage which is predicted by the TRO theory.

If the period variation is caused by a conservative mass transfer, then using
the well-known equation
\begin{equation}
\frac{\dot{P}}{P}=3\frac{\dot{M}_{2}}{M_{2}}(\frac{M_{2}}{M_{1}}-1),
\end{equation}
the mass transfer rate is estimated to be
${\rm d}M_{2}/{\rm d}t=1.4(\pm0.5)\times10^{-8}$ M$_{\odot}\cdot$yr$^{-1}$.
However, the real mass transfer rate may be quite different from this value because
of the contribution of angular momentum loss (AML) process \citep{Rahunen81,QianS13}.
In addition, it is possible that the long-term increase may be only one part
of a long cyclic oscillation. More observations are needed to clarify the nature of
the period variation.

\normalem
\begin{acknowledgements}
This work was partly supported by the Chinese Natural Science Foundation
(Nos. 11133007 and 11325315). New CCD photometric observations of
V1799 Ori were obtained with the 1.0-m telescope at Yunnan Observatories.
\end{acknowledgements}

\bibliographystyle{raa}
\bibliography{bibtex}

\clearpage
\begin{appendix}
\section{}

\begin{longtable}{cccccccccccc}
\caption[]{The B band original data of V1799 Ori observed in 2012 (Hel. JD 2,456,200+).}\label{tabA1}\\
\hline
Hel. JD & $\Delta m$  & Hel. JD& $\Delta m$ &  Hel. JD & $\Delta m$  &  Hel. JD &  $\Delta m$ &  Hel. JD & $\Delta m$ \\
\hline\noalign{\smallskip}
\endfirsthead
\caption[]{(continued)}\\
\hline
Hel. JD & $\Delta m$  & Hel. JD& $\Delta m$ &  Hel. JD & $\Delta m$  &  Hel. JD &  $\Delta m$ &  Hel. JD & $\Delta m$ \\
\hline\noalign{\smallskip}
\endhead
\hline
\endfoot
  76.9944 &  .151 &  77.0629 &  .090 &  77.1314 &  .253 &  77.1998 &  .111 &  77.2693 &  .315 \\
  76.9958 &  .074 &  77.0643 &  .095 &  77.1328 &  .198 &  77.2012 &  .132 &  77.2707 &  .310 \\
  76.9972 &  .132 &  77.0657 &  .146 &  77.1342 &  .210 &  77.2026 &  .113 &  77.2721 &  .304 \\
  76.9986 &  .092 &  77.0671 &  .114 &  77.1356 &  .170 &  77.2040 &  .147 &  77.2735 &  .264 \\
  77.0000 &  .098 &  77.0685 &  .165 &  77.1370 &  .162 &  77.2054 &  .143 &  77.2749 &  .235 \\
  77.0014 &  .085 &  77.0699 &  .200 &  77.1384 &  .128 &  77.2060 &  .154 &  77.2763 &  .207 \\
  77.0028 &  .098 &  77.0713 &  .184 &  77.1398 &  .159 &  77.2074 &  .152 &  77.2777 &  .187 \\
  77.0042 &  .027 &  77.0727 &  .252 &  77.1403 &  .170 &  77.2088 &  .142 &  77.2791 &  .186 \\
  77.0056 &  .089 &  77.0741 &  .237 &  77.1417 &  .112 &  77.2102 &  .156 &  77.2805 &  .150 \\
  77.0070 &  .036 &  77.0746 &  .263 &  77.1431 &  .115 &  77.2116 &  .194 &  77.2819 &  .162 \\
  77.0084 &  .083 &  77.0760 &  .280 &  77.1445 &  .088 &  77.2130 &  .181 &  77.2824 &  .139 \\
  77.0090 &  .053 &  77.0774 &  .345 &  77.1459 &  .152 &  77.2144 &  .241 &  77.2838 &  .137 \\
  77.0104 &  .047 &  77.0788 &  .397 &  77.1473 &  .086 &  77.2158 &  .228 &  77.2862 &  .107 \\
  77.0118 &  .022 &  77.0802 &  .405 &  77.1487 &  .084 &  77.2172 &  .213 &  77.2876 &  .136 \\
  77.0132 &  .010 &  77.0816 &  .481 &  77.1501 &  .115 &  77.2186 &  .286 &  77.2890 &  .075 \\
  77.0146 &  .068 &  77.0830 &  .487 &  77.1515 &  .100 &  77.2191 &  .281 &  77.2904 &  .107 \\
  77.0160 &  .000 &  77.0844 &  .509 &  77.1529 &  .042 &  77.2205 &  .295 &  77.2918 &  .072 \\
  77.0174 &  .025 &  77.0858 &  .565 &  77.1534 &  .061 &  77.2219 &  .288 &  77.2932 &  .062 \\
  77.0188 &  .023 &  77.0872 &  .561 &  77.1548 &  .095 &  77.2233 &  .337 &  77.2946 &  .053 \\
  77.0202 & -.003 &  77.0878 &  .663 &  77.1562 &  .056 &  77.2247 &  .359 &  77.2960 &  .067 \\
  77.0216 & -.015 &  77.0892 &  .688 &  77.1576 &  .071 &  77.2261 &  .410 &  77.2974 &  .056 \\
  77.0221 & -.040 &  77.0906 &  .741 &  77.1590 &  .013 &  77.2285 &  .428 &  77.2988 &  .065 \\
  77.0235 & -.020 &  77.0920 &  .837 &  77.1604 &  .044 &  77.2299 &  .476 &  77.3002 &  .058 \\
  77.0249 & -.016 &  77.0934 &  .862 &  77.1618 &  .032 &  77.2313 &  .490 &  77.3007 &  .050 \\
  77.0263 & -.034 &  77.0948 &  .896 &  77.1632 &  .004 &  77.2327 &  .469 &  77.3021 &  .012 \\
  77.0277 & -.034 &  77.0962 &  .957 &  77.1646 &  .017 &  77.2341 &  .554 &  77.3035 & -.034 \\
  77.0291 & -.032 &  77.0976 &  .954 &  77.1660 & -.003 &  77.2355 &  .641 &  77.3049 &  .014 \\
  77.0305 & -.039 &  77.0990 & 1.016 &  77.1666 &  .016 &  77.2369 &  .662 &  77.3063 & -.036 \\
  77.0319 & -.013 &  77.1004 & 1.033 &  77.1680 &  .014 &  77.2383 &  .654 &  77.3077 &  .017 \\
  77.0333 &  .001 &  77.1009 & 1.028 &  77.1694 &  .012 &  77.2397 &  .759 &  77.3091 &  .002 \\
  77.0347 & -.074 &  77.1023 & 1.035 &  77.1708 &  .016 &  77.2411 &  .779 &  77.3105 &  .030 \\
  77.0352 &  .002 &  77.1037 & 1.011 &  77.1722 &  .023 &  77.2425 &  .790 &  77.3119 & -.057 \\
  77.0366 & -.005 &  77.1051 &  .991 &  77.1736 &  .028 &  77.2430 &  .896 &  77.3133 &  .014 \\
  77.0380 & -.028 &  77.1065 &  .991 &  77.1750 &  .027 &  77.2444 &  .863 &  77.3139 & -.002 \\
  77.0394 &  .017 &  77.1079 &  .915 &  77.1764 &  .046 &  77.2458 &  .921 &  77.3153 & -.014 \\
  77.0408 &  .002 &  77.1093 &  .844 &  77.1778 &  .016 &  77.2472 &  .870 &  77.3167 & -.023 \\
  77.0422 & -.033 &  77.1107 &  .778 &  77.1792 &  .008 &  77.2486 &  .857 &  77.3181 &  .005 \\
  77.0436 & -.020 &  77.1121 &  .685 &  77.1797 &  .006 &  77.2500 &  .743 &  77.3195 & -.023 \\
  77.0450 & -.047 &  77.1135 &  .685 &  77.1811 &  .021 &  77.2514 &  .742 &  77.3209 & -.010 \\
  77.0464 &  .017 &  77.1140 &  .659 &  77.1825 &  .040 &  77.2528 &  .778 &  77.3223 & -.006 \\
  77.0478 &  .002 &  77.1154 &  .631 &  77.1839 &  .020 &  77.2542 &  .667 &  77.3237 & -.044 \\
  77.0484 & -.021 &  77.1168 &  .539 &  77.1853 &  .041 &  77.2556 &  .676 &  77.3251 & -.025 \\
  77.0498 & -.035 &  77.1182 &  .563 &  77.1867 &  .056 &  77.2561 &  .600 &  77.3265 & -.021 \\
  77.0512 &  .023 &  77.1196 &  .472 &  77.1881 &  .074 &  77.2575 &  .582 &  77.3270 & -.009 \\
  77.0526 &  .025 &  77.1210 &  .451 &  77.1895 &  .075 &  77.2589 &  .562 &  77.3284 & -.054 \\
  77.0540 &  .059 &  77.1224 &  .441 &  77.1909 &  .108 &  77.2603 &  .548 &  77.3298 & -.065 \\
  77.0554 &  .044 &  77.1238 &  .384 &  77.1923 &  .103 &  77.2617 &  .442 &  77.3312 & -.031 \\
  77.0568 &  .070 &  77.1252 &  .309 &  77.1928 &  .105 &  77.2631 &  .438 &  77.3326 & -.009 \\
  77.0582 &  .025 &  77.1266 &  .292 &  77.1942 &  .114 &  77.2645 &  .397 &  77.3340 &  .005 \\
  77.0596 &  .068 &  77.1272 &  .290 &  77.1956 &  .101 &  77.2659 &  .390 &  77.3354 &  .057 \\
  77.0610 &  .081 &  77.1286 &  .284 &  77.1970 &  .112 &  77.2673 &  .380 &  77.3368 &  .024 \\
  77.0615 &  .059 &  77.1300 &  .230 &  77.1984 &  .114 &  77.2687 &  .341 &                  \\
\noalign{\smallskip}\hline
\end{longtable}

\begin{longtable}{cccccccccc}
\caption[]{The V band original data of V1799 Ori observed in 2012 (Hel. JD 2,456,200+).}\\
\hline
Hel. JD & $\Delta m$  & Hel. JD& $\Delta m$ &  Hel. JD & $\Delta m$  &  Hel. JD &  $\Delta m$ &  Hel. JD & $\Delta m$ \\
\hline\noalign{\smallskip}
\endfirsthead
\caption[]{(continued)}\\
\hline
Hel. JD & $\Delta m$  & Hel. JD& $\Delta m$ &  Hel. JD & $\Delta m$  &  Hel. JD &  $\Delta m$ &  Hel. JD & $\Delta m$ \\
\hline\noalign{\smallskip}
\endhead
\hline
\endfoot
  76.9949 &   .004 &  77.0634 &   .017 &  77.1318 &   .149 &  77.2003 &   .007 &  77.2697 &   .234 \\
  76.9963 &   .063 &  77.0648 &   .051 &  77.1332 &   .143 &  77.2017 &   .015 &  77.2711 &   .201 \\
  76.9977 &   .012 &  77.0662 &   .066 &  77.1346 &   .083 &  77.2031 &   .020 &  77.2725 &   .168 \\
  76.9991 &   .023 &  77.0676 &   .064 &  77.1360 &   .086 &  77.2045 &   .036 &  77.2739 &   .145 \\
  77.0005 &   .017 &  77.0690 &   .091 &  77.1374 &   .069 &  77.2064 &   .026 &  77.2753 &   .137 \\
  77.0019 &  -.050 &  77.0704 &   .116 &  77.1388 &   .051 &  77.2078 &   .042 &  77.2767 &   .138 \\
  77.0033 &  -.035 &  77.0718 &   .133 &  77.1408 &   .031 &  77.2092 &   .068 &  77.2781 &   .114 \\
  77.0047 &  -.020 &  77.0732 &   .135 &  77.1422 &   .026 &  77.2106 &   .081 &  77.2795 &   .067 \\
  77.0061 &  -.017 &  77.0751 &   .223 &  77.1436 &   .035 &  77.2120 &   .092 &  77.2809 &   .067 \\
  77.0075 &  -.058 &  77.0765 &   .241 &  77.1450 &   .017 &  77.2134 &   .087 &  77.2829 &   .024 \\
  77.0094 &  -.023 &  77.0779 &   .228 &  77.1464 &  -.003 &  77.2149 &   .110 &  77.2843 &   .039 \\
  77.0108 &  -.055 &  77.0793 &   .294 &  77.1478 &  -.008 &  77.2163 &   .129 &  77.2867 &   .028 \\
  77.0122 &  -.040 &  77.0807 &   .339 &  77.1492 &  -.021 &  77.2177 &   .127 &  77.2881 &   .003 \\
  77.0136 &  -.059 &  77.0821 &   .366 &  77.1506 &  -.029 &  77.2196 &   .153 &  77.2895 &  -.001 \\
  77.0150 &  -.033 &  77.0835 &   .395 &  77.1520 &  -.037 &  77.2210 &   .183 &  77.2909 &   .005 \\
  77.0164 &  -.086 &  77.0849 &   .445 &  77.1539 &  -.024 &  77.2224 &   .209 &  77.2923 &  -.014 \\
  77.0178 &  -.049 &  77.0863 &   .474 &  77.1553 &  -.012 &  77.2238 &   .229 &  77.2937 &  -.020 \\
  77.0192 &  -.057 &  77.0882 &   .560 &  77.1567 &  -.046 &  77.2252 &   .252 &  77.2951 &  -.023 \\
  77.0206 &  -.095 &  77.0896 &   .609 &  77.1581 &  -.060 &  77.2266 &   .268 &  77.2965 &  -.035 \\
  77.0226 &  -.118 &  77.0910 &   .653 &  77.1595 &  -.041 &  77.2289 &   .353 &  77.2979 &  -.049 \\
  77.0240 &  -.099 &  77.0924 &   .742 &  77.1609 &  -.044 &  77.2303 &   .371 &  77.2993 &  -.072 \\
  77.0254 &  -.108 &  77.0938 &   .759 &  77.1623 &  -.063 &  77.2317 &   .408 &  77.3012 &  -.059 \\
  77.0268 &  -.117 &  77.0952 &   .795 &  77.1637 &  -.069 &  77.2331 &   .439 &  77.3026 &  -.087 \\
  77.0282 &  -.092 &  77.0966 &   .822 &  77.1651 &  -.070 &  77.2345 &   .465 &  77.3040 &  -.073 \\
  77.0296 &  -.091 &  77.0980 &   .900 &  77.1670 &  -.065 &  77.2359 &   .509 &  77.3054 &  -.086 \\
  77.0310 &  -.117 &  77.0994 &   .914 &  77.1684 &  -.070 &  77.2373 &   .548 &  77.3068 &  -.075 \\
  77.0324 &  -.099 &  77.1014 &   .891 &  77.1698 &  -.059 &  77.2387 &   .578 &  77.3082 &  -.089 \\
  77.0338 &  -.102 &  77.1028 &   .919 &  77.1712 &  -.082 &  77.2401 &   .628 &  77.3096 &  -.104 \\
  77.0357 &  -.087 &  77.1042 &   .949 &  77.1726 &  -.064 &  77.2415 &   .692 &  77.3110 &  -.113 \\
  77.0371 &  -.092 &  77.1056 &   .887 &  77.1740 &  -.045 &  77.2435 &   .759 &  77.3124 &  -.115 \\
  77.0385 &  -.107 &  77.1070 &   .826 &  77.1754 &  -.066 &  77.2449 &   .765 &  77.3143 &  -.132 \\
  77.0399 &  -.107 &  77.1084 &   .761 &  77.1768 &  -.061 &  77.2463 &   .743 &  77.3157 &  -.117 \\
  77.0413 &  -.055 &  77.1098 &   .715 &  77.1782 &  -.069 &  77.2477 &   .753 &  77.3171 &  -.114 \\
  77.0427 &  -.115 &  77.1112 &   .647 &  77.1802 &  -.069 &  77.2491 &   .696 &  77.3185 &  -.099 \\
  77.0441 &  -.094 &  77.1126 &   .615 &  77.1816 &  -.046 &  77.2505 &   .713 &  77.3199 &  -.103 \\
  77.0455 &  -.086 &  77.1145 &   .544 &  77.1830 &  -.068 &  77.2519 &   .657 &  77.3213 &  -.090 \\
  77.0469 &  -.075 &  77.1159 &   .495 &  77.1844 &  -.057 &  77.2533 &   .628 &  77.3227 &  -.119 \\
  77.0488 &  -.078 &  77.1173 &   .451 &  77.1858 &  -.046 &  77.2547 &   .564 &  77.3242 &  -.102 \\
  77.0502 &  -.062 &  77.1187 &   .422 &  77.1872 &  -.041 &  77.2566 &   .500 &  77.3256 &  -.106 \\
  77.0516 &  -.059 &  77.1201 &   .380 &  77.1886 &  -.065 &  77.2580 &   .505 &  77.3275 &  -.113 \\
  77.0530 &  -.037 &  77.1215 &   .330 &  77.1900 &  -.031 &  77.2594 &   .443 &  77.3289 &  -.081 \\
  77.0544 &  -.028 &  77.1229 &   .307 &  77.1914 &  -.027 &  77.2608 &   .411 &  77.3303 &  -.101 \\
  77.0558 &  -.020 &  77.1243 &   .285 &  77.1933 &  -.014 &  77.2622 &   .373 &  77.3317 &  -.091 \\
  77.0572 &   .008 &  77.1257 &   .248 &  77.1947 &  -.011 &  77.2636 &   .349 &  77.3331 &  -.080 \\
  77.0586 &  -.003 &  77.1276 &   .198 &  77.1961 &  -.007 &  77.2650 &   .316 &  77.3345 &  -.085 \\
  77.0600 &  -.003 &  77.1290 &   .186 &  77.1975 &  -.014 &  77.2664 &   .271 &  77.3359 &  -.080 \\
  77.0620 &   .002 &  77.1304 &   .159 &  77.1989 &   .006 &  77.2678 &   .262 &  77.3373 &  -.046 \\
\noalign{\smallskip}\hline
\end{longtable}

\begin{longtable}{cccccccccc}
\caption[]{The R band original data of V1799 Ori observed in 2012 (Hel. JD 2,456,200+).}\\
\hline
Hel. JD & $\Delta m$  & Hel. JD& $\Delta m$ &  Hel. JD & $\Delta m$  &  Hel. JD &  $\Delta m$ &  Hel. JD & $\Delta m$ \\
\hline\noalign{\smallskip}
\endfirsthead
\caption[]{(continued)}\\
\hline
Hel. JD & $\Delta m$  & Hel. JD& $\Delta m$ &  Hel. JD & $\Delta m$  &  Hel. JD &  $\Delta m$ &  Hel. JD & $\Delta m$ \\
\hline\noalign{\smallskip}
\endhead
\hline
\endfoot
  76.9952 &  -.069 &  77.0637 &  -.059 &  77.1322 &   .047 &  77.2006 &  -.096 &  77.2701 &   .118 \\
  76.9966 &  -.084 &  77.0651 &  -.048 &  77.1336 &   .017 &  77.2020 &  -.106 &  77.2715 &   .054 \\
  76.9980 &  -.082 &  77.0665 &  -.044 &  77.1350 &  -.012 &  77.2034 &  -.097 &  77.2729 &   .047 \\
  76.9994 &  -.078 &  77.0679 &  -.019 &  77.1364 &  -.010 &  77.2048 &  -.076 &  77.2743 &   .033 \\
  77.0008 &  -.132 &  77.0693 &   .019 &  77.1378 &  -.022 &  77.2068 &  -.046 &  77.2757 &   .044 \\
  77.0022 &  -.121 &  77.0707 &   .029 &  77.1392 &  -.049 &  77.2082 &  -.043 &  77.2771 &   .016 \\
  77.0036 &  -.122 &  77.0721 &   .054 &  77.1411 &  -.065 &  77.2096 &  -.024 &  77.2785 &  -.017 \\
  77.0050 &  -.133 &  77.0735 &   .072 &  77.1425 &  -.076 &  77.2110 &  -.009 &  77.2799 &  -.037 \\
  77.0064 &  -.156 &  77.0754 &   .115 &  77.1439 &  -.084 &  77.2124 &  -.011 &  77.2813 &  -.050 \\
  77.0078 &  -.140 &  77.0768 &   .125 &  77.1453 &  -.082 &  77.2138 &   .016 &  77.2832 &  -.070 \\
  77.0098 &  -.124 &  77.0782 &   .171 &  77.1467 &  -.098 &  77.2152 &   .019 &  77.2846 &  -.066 \\
  77.0112 &  -.158 &  77.0796 &   .184 &  77.1481 &  -.084 &  77.2166 &   .025 &  77.2870 &  -.088 \\
  77.0126 &  -.154 &  77.0810 &   .225 &  77.1495 &  -.124 &  77.2180 &   .036 &  77.2884 &  -.082 \\
  77.0140 &  -.134 &  77.0824 &   .279 &  77.1509 &  -.117 &  77.2199 &   .077 &  77.2898 &  -.110 \\
  77.0154 &  -.148 &  77.0838 &   .295 &  77.1523 &  -.117 &  77.2213 &   .090 &  77.2912 &  -.100 \\
  77.0168 &  -.175 &  77.0852 &   .329 &  77.1542 &  -.124 &  77.2227 &   .114 &  77.2926 &  -.125 \\
  77.0182 &  -.151 &  77.0866 &   .397 &  77.1556 &  -.113 &  77.2241 &   .137 &  77.2940 &  -.117 \\
  77.0196 &  -.152 &  77.0886 &   .454 &  77.1570 &  -.127 &  77.2255 &   .170 &  77.2954 &  -.125 \\
  77.0210 &  -.178 &  77.0900 &   .463 &  77.1584 &  -.143 &  77.2269 &   .202 &  77.2968 &  -.132 \\
  77.0229 &  -.192 &  77.0914 &   .570 &  77.1598 &  -.141 &  77.2293 &   .233 &  77.2982 &  -.141 \\
  77.0243 &  -.191 &  77.0928 &   .591 &  77.1612 &  -.153 &  77.2307 &   .279 &  77.2996 &  -.159 \\
  77.0257 &  -.183 &  77.0942 &   .631 &  77.1626 &  -.151 &  77.2321 &   .299 &  77.3015 &  -.150 \\
  77.0271 &  -.179 &  77.0956 &   .690 &  77.1640 &  -.164 &  77.2335 &   .351 &  77.3029 &  -.167 \\
  77.0285 &  -.210 &  77.0970 &   .712 &  77.1654 &  -.158 &  77.2349 &   .355 &  77.3043 &  -.158 \\
  77.0299 &  -.195 &  77.0984 &   .760 &  77.1674 &  -.166 &  77.2363 &   .425 &  77.3057 &  -.169 \\
  77.0313 &  -.182 &  77.0998 &   .756 &  77.1688 &  -.171 &  77.2377 &   .448 &  77.3071 &  -.178 \\
  77.0327 &  -.201 &  77.1017 &   .772 &  77.1702 &  -.175 &  77.2391 &   .501 &  77.3085 &  -.168 \\
  77.0341 &  -.198 &  77.1031 &   .788 &  77.1716 &  -.163 &  77.2405 &   .544 &  77.3099 &  -.174 \\
  77.0360 &  -.187 &  77.1045 &   .774 &  77.1730 &  -.152 &  77.2419 &   .567 &  77.3113 &  -.174 \\
  77.0374 &  -.170 &  77.1059 &   .733 &  77.1744 &  -.174 &  77.2438 &   .577 &  77.3127 &  -.182 \\
  77.0388 &  -.151 &  77.1073 &   .678 &  77.1758 &  -.158 &  77.2452 &   .630 &  77.3147 &  -.169 \\
  77.0402 &  -.144 &  77.1087 &   .636 &  77.1772 &  -.153 &  77.2466 &   .625 &  77.3161 &  -.184 \\
  77.0416 &  -.177 &  77.1101 &   .580 &  77.1786 &  -.171 &  77.2480 &   .618 &  77.3175 &  -.183 \\
  77.0430 &  -.200 &  77.1115 &   .524 &  77.1805 &  -.164 &  77.2494 &   .610 &  77.3189 &  -.184 \\
  77.0444 &  -.168 &  77.1129 &   .464 &  77.1819 &  -.154 &  77.2508 &   .592 &  77.3203 &  -.176 \\
  77.0458 &  -.175 &  77.1148 &   .417 &  77.1833 &  -.159 &  77.2522 &   .515 &  77.3217 &  -.193 \\
  77.0472 &  -.145 &  77.1162 &   .377 &  77.1847 &  -.158 &  77.2536 &   .486 &  77.3231 &  -.179 \\
  77.0492 &  -.149 &  77.1176 &   .325 &  77.1861 &  -.152 &  77.2550 &   .442 &  77.3245 &  -.196 \\
  77.0506 &  -.129 &  77.1190 &   .299 &  77.1875 &  -.144 &  77.2569 &   .384 &  77.3259 &  -.180 \\
  77.0520 &  -.176 &  77.1204 &   .261 &  77.1889 &  -.135 &  77.2583 &   .379 &  77.3278 &  -.158 \\
  77.0534 &  -.119 &  77.1218 &   .207 &  77.1903 &  -.135 &  77.2597 &   .325 &  77.3292 &  -.165 \\
  77.0548 &  -.110 &  77.1232 &   .165 &  77.1917 &  -.130 &  77.2611 &   .288 &  77.3306 &  -.164 \\
  77.0562 &  -.091 &  77.1246 &   .154 &  77.1936 &  -.108 &  77.2625 &   .284 &  77.3320 &  -.160 \\
  77.0576 &  -.083 &  77.1260 &   .126 &  77.1950 &  -.115 &  77.2639 &   .230 &  77.3334 &  -.162 \\
  77.0590 &  -.100 &  77.1280 &   .100 &  77.1964 &  -.132 &  77.2653 &   .206 &  77.3348 &  -.157 \\
  77.0604 &  -.085 &  77.1294 &   .064 &  77.1978 &  -.109 &  77.2667 &   .212 &  77.3362 &  -.156 \\
  77.0623 &  -.067 &  77.1308 &   .057 &  77.1992 &  -.097 &  77.2681 &   .144 &  77.3376 &  -.133 \\
\noalign{\smallskip}\hline
\end{longtable}

\begin{longtable}{cccccccccc}
\caption[]{The I band original data of V1799 Ori observed in 2012 (Hel. JD 2,456,200+).}\label{tabA4}\\
\hline
Hel. JD & $\Delta m$  & Hel. JD& $\Delta m$ &  Hel. JD & $\Delta m$  &  Hel. JD &  $\Delta m$ &  Hel. JD & $\Delta m$ \\
\hline\noalign{\smallskip}
\endfirsthead
\caption[]{(continued)}\\
\hline
Hel. JD & $\Delta m$  & Hel. JD& $\Delta m$ &  Hel. JD & $\Delta m$  &  Hel. JD &  $\Delta m$ &  Hel. JD & $\Delta m$ \\
\hline\noalign{\smallskip}
\endhead
\hline
\endfoot
  76.9954 &  -.131 &  77.0639 &  -.130 &  77.1324 &  -.065 &  77.2009 &  -.160 &  77.2703 &   .024 \\
  76.9968 &  -.133 &  77.0653 &  -.116 &  77.1338 &  -.057 &  77.2023 &  -.157 &  77.2717 &   .022 \\
  76.9982 &  -.147 &  77.0667 &  -.100 &  77.1352 &  -.092 &  77.2037 &  -.167 &  77.2731 &   .008 \\
  76.9997 &  -.165 &  77.0681 &  -.105 &  77.1366 &  -.083 &  77.2051 &  -.158 &  77.2745 &  -.026 \\
  77.0011 &  -.156 &  77.0695 &  -.073 &  77.1380 &  -.098 &  77.2070 &  -.130 &  77.2759 &  -.041 \\
  77.0025 &  -.187 &  77.0709 &  -.046 &  77.1394 &  -.121 &  77.2084 &  -.129 &  77.2773 &  -.027 \\
  77.0039 &  -.188 &  77.0723 &  -.011 &  77.1413 &  -.147 &  77.2098 &  -.102 &  77.2787 &  -.064 \\
  77.0053 &  -.176 &  77.0737 &  -.014 &  77.1427 &  -.156 &  77.2112 &  -.089 &  77.2801 &  -.086 \\
  77.0067 &  -.170 &  77.0756 &   .041 &  77.1441 &  -.125 &  77.2126 &  -.059 &  77.2815 &  -.108 \\
  77.0081 &  -.162 &  77.0770 &   .056 &  77.1455 &  -.151 &  77.2140 &  -.079 &  77.2834 &  -.138 \\
  77.0100 &  -.225 &  77.0784 &   .104 &  77.1469 &  -.193 &  77.2154 &  -.057 &  77.2848 &  -.127 \\
  77.0114 &  -.198 &  77.0798 &   .127 &  77.1483 &  -.177 &  77.2168 &  -.041 &  77.2872 &  -.143 \\
  77.0128 &  -.181 &  77.0812 &   .175 &  77.1497 &  -.186 &  77.2182 &  -.018 &  77.2886 &  -.164 \\
  77.0142 &  -.194 &  77.0827 &   .165 &  77.1511 &  -.173 &  77.2201 &   .023 &  77.2900 &  -.155 \\
  77.0156 &  -.212 &  77.0841 &   .235 &  77.1525 &  -.186 &  77.2215 &   .038 &  77.2914 &  -.170 \\
  77.0170 &  -.216 &  77.0855 &   .262 &  77.1545 &  -.162 &  77.2229 &   .056 &  77.2928 &  -.165 \\
  77.0184 &  -.210 &  77.0869 &   .298 &  77.1559 &  -.184 &  77.2243 &   .081 &  77.2942 &  -.179 \\
  77.0198 &  -.222 &  77.0888 &   .341 &  77.1573 &  -.198 &  77.2257 &   .105 &  77.2956 &  -.182 \\
  77.0212 &  -.247 &  77.0902 &   .387 &  77.1587 &  -.197 &  77.2271 &   .118 &  77.2970 &  -.187 \\
  77.0231 &  -.243 &  77.0916 &   .456 &  77.1601 &  -.235 &  77.2295 &   .178 &  77.2984 &  -.212 \\
  77.0245 &  -.271 &  77.0930 &   .512 &  77.1615 &  -.220 &  77.2309 &   .210 &  77.2998 &  -.196 \\
  77.0259 &  -.269 &  77.0944 &   .561 &  77.1629 &  -.193 &  77.2323 &   .237 &  77.3018 &  -.209 \\
  77.0273 &  -.258 &  77.0958 &   .618 &  77.1643 &  -.221 &  77.2337 &   .280 &  77.3032 &  -.204 \\
  77.0287 &  -.247 &  77.0972 &   .660 &  77.1657 &  -.223 &  77.2351 &   .315 &  77.3046 &  -.229 \\
  77.0301 &  -.265 &  77.0986 &   .675 &  77.1676 &  -.209 &  77.2365 &   .343 &  77.3060 &  -.234 \\
  77.0315 &  -.237 &  77.1000 &   .665 &  77.1690 &  -.210 &  77.2379 &   .407 &  77.3074 &  -.227 \\
  77.0329 &  -.252 &  77.1019 &   .655 &  77.1704 &  -.227 &  77.2393 &   .422 &  77.3088 &  -.229 \\
  77.0343 &  -.258 &  77.1033 &   .651 &  77.1718 &  -.201 &  77.2407 &   .481 &  77.3102 &  -.232 \\
  77.0362 &  -.233 &  77.1047 &   .668 &  77.1732 &  -.219 &  77.2421 &   .499 &  77.3116 &  -.236 \\
  77.0376 &  -.235 &  77.1061 &   .629 &  77.1746 &  -.217 &  77.2440 &   .551 &  77.3130 &  -.233 \\
  77.0390 &  -.246 &  77.1075 &   .570 &  77.1760 &  -.220 &  77.2454 &   .569 &  77.3149 &  -.247 \\
  77.0404 &  -.257 &  77.1089 &   .513 &  77.1774 &  -.221 &  77.2468 &   .538 &  77.3163 &  -.249 \\
  77.0418 &  -.228 &  77.1103 &   .482 &  77.1788 &  -.225 &  77.2482 &   .541 &  77.3177 &  -.256 \\
  77.0432 &  -.226 &  77.1117 &   .414 &  77.1807 &  -.213 &  77.2496 &   .529 &  77.3191 &  -.235 \\
  77.0446 &  -.235 &  77.1131 &   .358 &  77.1821 &  -.239 &  77.2510 &   .497 &  77.3205 &  -.240 \\
  77.0461 &  -.231 &  77.1151 &   .309 &  77.1835 &  -.209 &  77.2524 &   .455 &  77.3219 &  -.237 \\
  77.0475 &  -.228 &  77.1165 &   .287 &  77.1849 &  -.211 &  77.2538 &   .410 &  77.3233 &  -.226 \\
  77.0494 &  -.224 &  77.1179 &   .234 &  77.1863 &  -.220 &  77.2552 &   .388 &  77.3247 &  -.236 \\
  77.0508 &  -.198 &  77.1193 &   .187 &  77.1877 &  -.202 &  77.2571 &   .324 &  77.3261 &  -.236 \\
  77.0522 &  -.244 &  77.1207 &   .160 &  77.1891 &  -.204 &  77.2585 &   .298 &  77.3280 &  -.214 \\
  77.0536 &  -.203 &  77.1221 &   .153 &  77.1905 &  -.201 &  77.2599 &   .272 &  77.3294 &  -.227 \\
  77.0550 &  -.183 &  77.1235 &   .095 &  77.1919 &  -.177 &  77.2613 &   .234 &  77.3308 &  -.226 \\
  77.0564 &  -.178 &  77.1249 &   .076 &  77.1939 &  -.179 &  77.2627 &   .210 &  77.3322 &  -.212 \\
  77.0578 &  -.156 &  77.1263 &   .051 &  77.1953 &  -.176 &  77.2641 &   .159 &  77.3336 &  -.207 \\
  77.0592 &  -.143 &  77.1282 &   .011 &  77.1967 &  -.171 &  77.2655 &   .123 &  77.3350 &  -.211 \\
  77.0606 &  -.170 &  77.1296 &   .012 &  77.1981 &  -.184 &  77.2669 &   .111 &  77.3364 &  -.200 \\
  77.0625 &  -.133 &  77.1310 &  -.029 &  77.1995 &  -.162 &  77.2683 &   .044 &  77.3378 &  -.193 \\
\noalign{\smallskip}\hline
\end{longtable}

\end{appendix}

\end{document}